\documentclass{PoS}
\usepackage{graphicx}
\usepackage{float}
\usepackage{wrapfig}

\title{An update on the developmental status of the Spin-Light Polarimeter for the Electron Ion Collider}

\ShortTitle{A Novel Spin-Light Polarimeter}

\author{\speaker{Prajwal Mohanmurthy}\thanks{This work is supported by a JSA Fellowship and a BNL Electron Ion Collider R \& D grant.}\\
        Laboratory for Nuclear Science and Department of Physics \\Massachusetts Institute of Technology, Cambridge, MA 02139\\
        E-mail: \email{prajwal@mit.edu}}

\author{Dipangkar Dutta\\
       Department of Physics and Astronomy\\Mississippi State University, MS 39762-5167\\
        E-mail: \email{d.dutta@msstate.edu}}

\abstract{
Precision experiments in the parity violating electron scattering (PVES) sector is one the leading methods to probe physics beyond the standard model (SM). A large part of the physics program being envisioned for future facilities such as the Electron Ion Collider (EIC) includes searching for physics beyond SM. Here, we present a novel technique which uses spacial asymmetry of synchrotron radiation produced by an electron beam passing through a wiggler magnet to trace the changes in beam polarization. Such a relative polarimeter could be vital if the goal of $< 0.5\%$ polarimetry is to be achieved at EIC. In this paper, we update the discussion on the development of this technique supported by a Geant4 simulation. The polarimeter apparatus along with the underlying basic ideas are briefly introduced. As a part of the simulation, the effects of electron beam current and beam energy were studied which were found to be manageable over a wide range of electron beam energies and beam currents. It was found that such a relative polarimeter works best in the $4-20$ GeV regime.
}
\FullConference{XVth International Workshop on Polarized Sources, Targets, and Polarimetry,\\
		September 9-13, 2013\\
		Charlottesville, Virginia, USA}

\begin{document}
\vspace{-10mm}
\section{Introduction}
\vspace{-3mm}
\begin{wrapfigure}{h!}{0.5\textwidth}
\vspace{-15mm}
\centering
    \includegraphics[scale=.14]{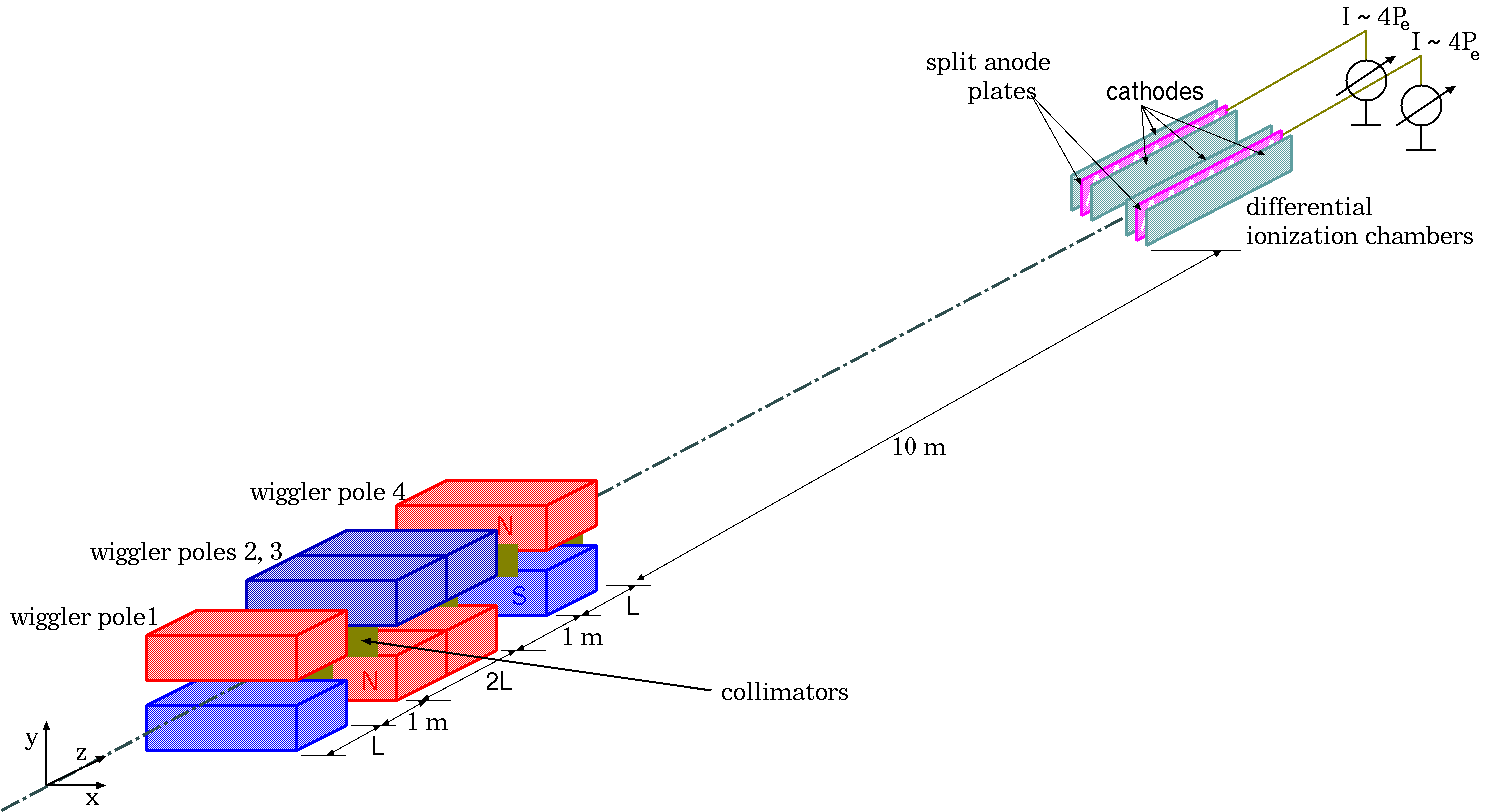}
\caption[]{Schematic diagram of the Spin-Light polarimeter apparatus.}
\label{fig1}
\vspace{-3mm}
\end{wrapfigure}

PVES experiments of the future demand a high degree of precision in polarimetry which can only be achieved if more than one polarimeter is used in series, therefore new polarimetry techniques, besides the conventional Compton and M\o ller polarimeters, are required. Karabekov and Rossmanith had already come up with the idea of using synchrotron radiation (SR) to measure beam polarization \cite{[1]}. Improving on the original proposal, the technique of the Spin-Light polarimeter here takes a step towards building a full fledged precision device. However, the power output of the wiggler magnets and resolution required to measure the spacial asymmetry of the synchrotron radiation produced by high energy beams constrains the electron beam energies at which such a polarimeter works best to below $20 GeV$. A lower limit of $4GeV$ on electron beam energies is imposed by the fact that at low energies, the spacial asymmetry of synchrotron radiation reduces making it very difficult to measure the polarization.
\vspace{-1mm}

\section{Synchrotron Light}
\vspace{-3mm}
\begin{wrapfigure}{h!}{0.5\textwidth}
\vspace{-20mm}
\centering
    \includegraphics[scale=.25]{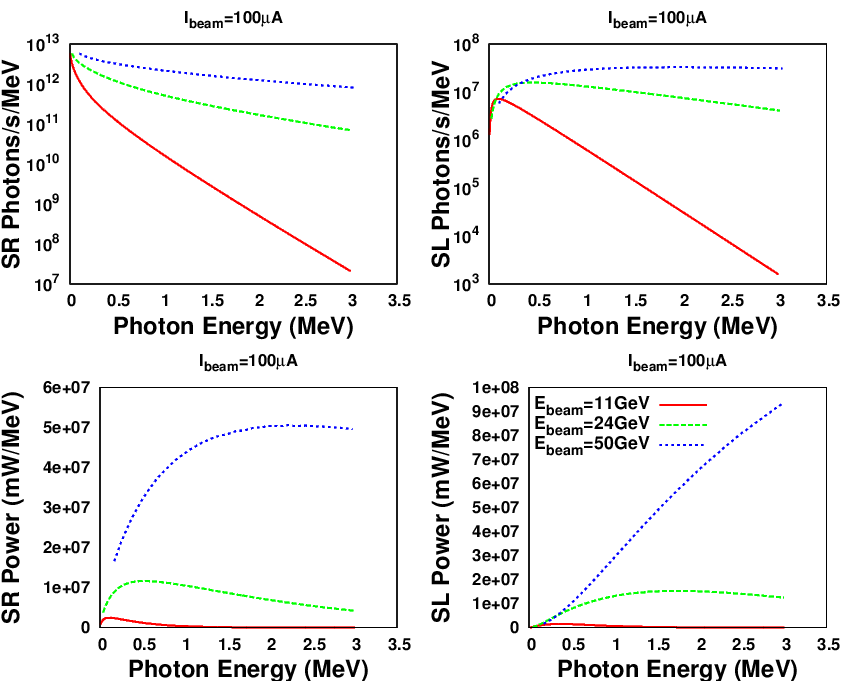}
\caption[]{Figures showing SR \& SL total number and total power spectra emitted by a wiggler magnet $B_{wig}=4T$ carrying an electron beam current of $100\mu A$ (Clockwise from Top-Left): A.,B. SR and SL photons per $MeV$ as a function of electron beam energy.;C.,D. SL and SR power per $MeV$  as a function of electron beam energy.}
\label{fig2}
\vspace{-3mm}
\end{wrapfigure}

Sokolov et. al. \cite{[2]} give the spin dependence of synchrotron radiation, produced by an electron passing through a magnetic field, in terms of convenient parameters including the SR photon's vertical opening angle - $\psi$ (angle between the
momentum component of the SR photon in the $y - z$ plane and the $z$ axis) and its longitudinal polarization - $\zeta$.  One can obtain the power spectra ($P_{\gamma}$) by multiplying the number of SR photons (Eq~$(2.1)$) with their corresponding energy bin. The difference of the integrated SR power spectra above ({\it i.e.} $0 \leq \psi \leq \pi/2$) and below ({\it i.e.} $-\pi/2 \leq \psi \leq 0$) the trajectory of the electrons turns out to be directly proportional to the longitudinal polarization of the electrons, with an offset term ($P_{Unpol}$) arising due to non-zero, but spin-independent, integrated (over all energies) power for a spin-averaged electron passing through a magnetic field. In this paper, the quantity in Eq. ($2.1$) is referred to as \emph{spin-light} (SL) \cite{[6]}.
\begin{eqnarray}
\vspace{-5mm}
 N_{\gamma} & = & \frac{9 n_e}{16 \pi^3}\frac{e^2}{cm_eR^2}\gamma^4\int_{0}^{\infty}dy \frac{y^2}{(1+\xi y)^4}\oint d\Omega (1 + \alpha ^2)^2 \times \nonumber\\
& & \left[K^2_{2/3}(z) + \frac{\alpha^2}{1 + \alpha^2}K^2_{1/3}(z) 
+ \zeta \xi y \frac{\alpha}{\sqrt{1+\alpha^2}}K_{1/3}(z)K_{2/3}(z)\right]
\end{eqnarray}
\vspace{-5mm}
\begin{eqnarray}
\Delta P_{\gamma} = P_{Pol} - P_{Unpol} & = & - \zeta \xi P_{Clas} \int_0^{\infty}dy\frac{9\sqrt{3}}{8\pi}y^2 K_{1/3}(y)
\vspace{-5mm}
\end{eqnarray}

\begin{wrapfigure}{h!}{0.5\textwidth}
\vspace{-5mm}
\centering
    \includegraphics[scale=.25]{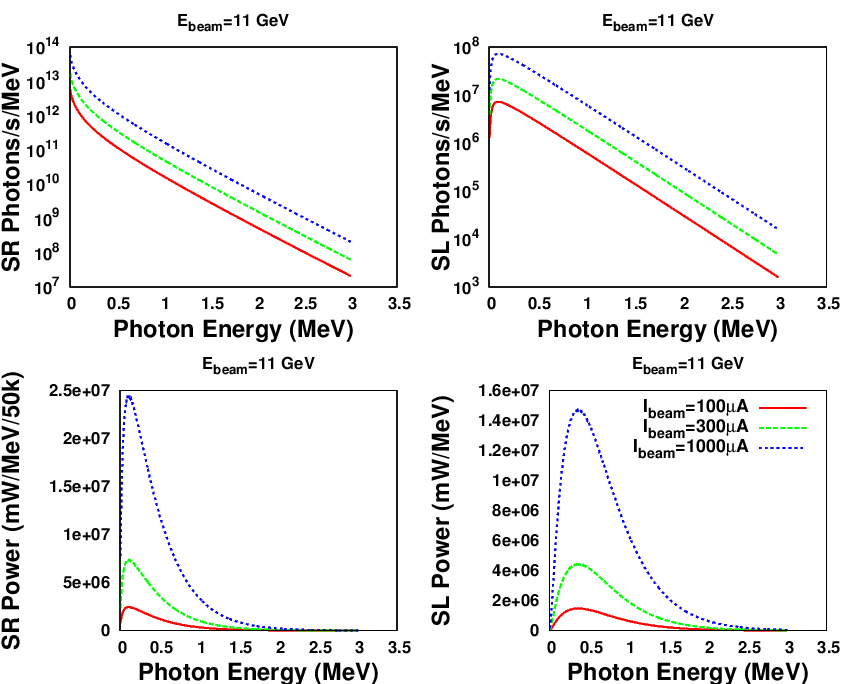}
\caption[]{Figures showing SR \& SL total number and total power spectra emitted by a wiggler magnet $B_{wig}=4T$ carrying an electron beam with an energy of $11GeV$ (Clockwise from Top-Left): A.,B. SR and SL photons per $MeV$ as a function of electron beam current.;C.,D. SL and SR power per $MeV$as a function of electron beam current.}
\label{fig3}
\vspace{-15mm}
\end{wrapfigure}

\noindent where $ \xi  = 3 B_{wig} / (2 B_c) $, $B_c$ being the magnetic field under the influence of which the entire kinetic energy of the electron is expelled as one SR photon, $y = \omega/\omega_c$, $K_n(x)$ are modified Bessel functions, $n_e$ is the number of electrons and, $z= \omega (1 + \alpha^2)^{3/2}/(2\omega_C)$, and $\alpha = \gamma \psi$. An asymmetry term, $A = \Delta N_{\gamma}/N_{\gamma}$ can then be defined from the above two equations, where $\Delta N_{\gamma}$ corresponds to $\Delta P_{\gamma}$.

\section{Spin-Light Polarimeter}
\vspace{-3mm}
A most basic layout of the spin-light polarimeter would include a 3-pole wiggler magnet, collimators and a split plane ionization chamber (IC) as illustrated in Figure~\ref{fig1}\cite{[3]}.


\subsection{Wiggler Magnets and Collimators}
\vspace{-1mm}
When an electron beam passes through a set of three wiggler (chicane like) magnets, where the central pole is twice as long as the poles on the extremes, the electron beam produces SR photons which contains the beam polarization information. All the individual magnetic poles would have equal field strength but the direction of the field of the central pole would be opposite to that of the poles on the extremes so that the electron beam direction ultimately remains unchanged.

\begin{wrapfigure}{h!}{0.45\textwidth}
\vspace{-4mm}
\centering
    \includegraphics[scale=.3]{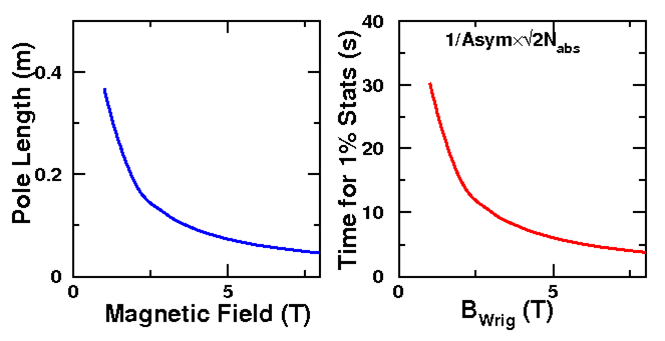}
\caption[]{(Left - Right): A. Pole length of the wiggler magnet required to give a SR - light cone with $1mrad$ spread as a function of wiggler magnetic field strength; B. Approximate time required to achieve $1\%$ statistical uncertainty as a function of wiggler magnetic field strength.}
\label{fig5}
\vspace{-3mm}
\end{wrapfigure}

\begin{wrapfigure}{h}{0.5\textwidth}
\vspace{-10mm}
\centering
    \includegraphics[scale=.2]{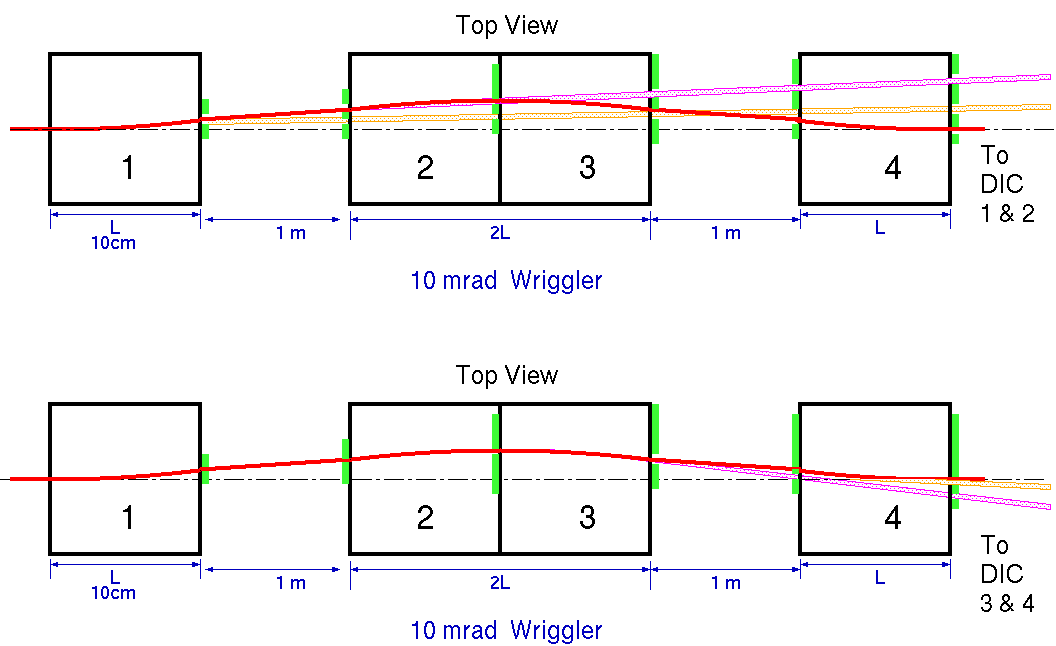}
\caption[]{(Top - Bottom): A. Schematic diagram showing the location of collimators (yellow strips) on the wiggler pole faces which guide the SR photons produced at the wiggler magnet to the beam left ionization chamber.; B. Schematic diagram showing the location of collimators on the wiggler pole faces which guide the SR photons produced at the wiggler magnet to the beam right ionization chamber. }
\label{fig5b}
\vspace{-5mm}
\end{wrapfigure}

The wiggler pole length, required to obtain a SR fan which has an angular spread of $1mrad$ in $\psi$, as a function of field strength could easily be calculated and is plotted in Figure~\ref{fig5}A. The magnetic field strength of $4T$ corresponding to a pole length of about $10cm$ was chosen owing to the availability of magnets with similar field strength. The effects of such wiggler magnets on the quality of electron beam was studied in Ref. \cite{[3]} and was reported to be negligible. Collimator slits in front of both the faces (faces through which the beam enters and exits the wiggler poles) of each wiggler magnet pole allows collimation of the SR beam, as illustrated in Figure~\ref{fig5b}, and directs them to two identical split-plane ionization chambers, one each on beam right and beam left positions down the beam-line from the wiggler magnets. With the help of collimators (yellow strips in Figure~\ref{fig5b}), one narrow beam of SR photons can be extracted from each wiggler pole, numbered 1 through 4. Of the four narrow beams extracted with the help of collimators, two of the narrow beams are directed to one ionization chamber while the other two narrow beams are directed to the other ionization chamber. 

Notice that the collimators truncate the SR fan produced at the wigglers in an angle-$\phi$ (angle between the momentum component of the SR photon in the $x - z$ plane and the $z$ axis), which is orthogonal to the angle-$\psi$, where SR photon distribution in $\psi$ contains the electron beam polarization information. Another important design consideration is the amount of time required to achieve statistically competent data which is inversely proportional to the product of asymmetry and the SR flux reaching the ionization chamber after collimation. In Figure~\ref{fig5}B, a plot of the amount of time required to obtain $1\%$ uncertainty in polarization measured as a function of magnetic field indicated that only about $10$ minutes are required if the wigglers were operated at a field strength of $4T$, a fairly plausible time scale for each measurement.


\subsection{Spin-Light Characteristics}
\vspace{-1mm}
\begin{wrapfigure}{h!}{0.55\textwidth}
\vspace{-20mm}
\centering
    \includegraphics[scale=.3]{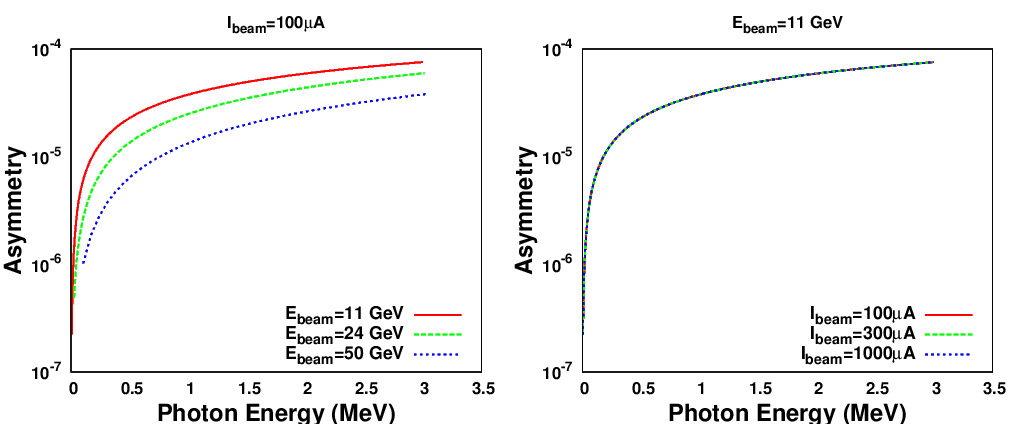}
\caption[]{(Left-Right): A. A plot of asymmetry in the SR light fan as a function of the SR photon energy for various electron beam energies and a fixed electron beam current of $100\mu A$.; B. A plot of asymmetry in the SR light fan as a function of the SR photon energy for various electron beam currents and a fixed electron beam energy of $11GeV$.}
\label{fig6}
\vspace{-3mm}
\end{wrapfigure}

Now that the parameters of the apparatus that generates SR photons are fixed, we can look at the charecteristics of SL photons in contrast to that of SR photons. Eq. ($2.1$) can be numerically integrated to plot the SR \& SL number spectra as a function of photon energy, while the SR \& SL power spectra could then be obtained from the corresponding number spectra as illustrated in Figures~\ref{fig2} \& \ref{fig3}. In Figures~\ref{fig2} \& \ref{fig3}, the plots have been made for various electron beam energies holding the electron beam current constant at $100\mu A$ and for various electron beam currents holding the electron beam energy constant at $11GeV$ respectively. Similarly, by numerically integrating Eq. ($2.2$), one could plot the asymmetry in the SR fan as illustrated in Figure~\ref{fig6}. An observation of interest here would be that while there is a slight drop in overall asymmetry with increase in electron beam energy, the asymmetry distribution effectively remains unchanged over a wide range of electron beam currents.
\begin{figure}[h!]
\vspace{-1mm}
\centering
    \includegraphics[scale=.375]{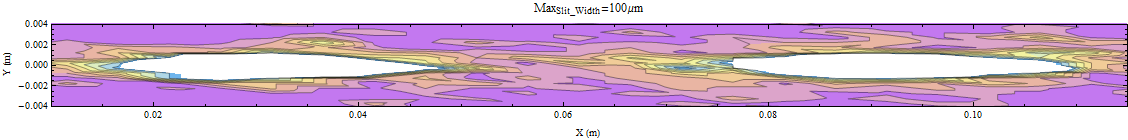}
\caption[]{2D projection of the collimated SR distribution in the ionization chamber as generated by the Geant4 simulation, the lighter areas have higher flux.}
\label{fig7}
\vspace{-3mm}
\end{figure}

\subsection{Ionization Chambers}
\begin{wrapfigure}{h!}{0.5\textwidth}
\vspace{-17mm}
\centering
    \includegraphics[scale=.5]{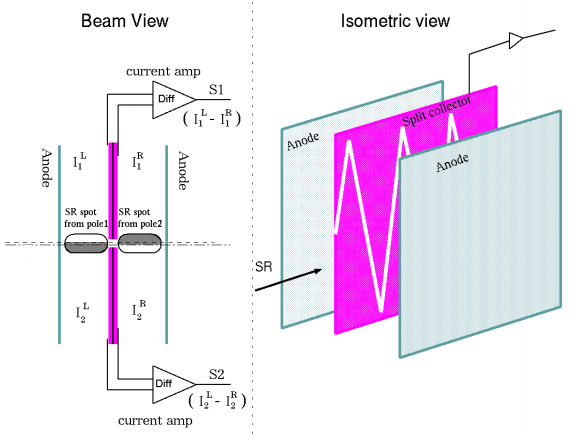}
\caption[]{(Left-Right) A. A schematic diagram of the two collimated narrow SR beams (ovals) incident on beam left split plane ionization chambers.; B. An isometric view of the split plane IC showing the back-gammon like central cathode sandwiched between two anodes.}
\label{fig7m}
\vspace{-3mm}
\end{wrapfigure}

The ionization chambers required for a spin-light polarimeter need to be sensitive to fine spacial asymmetry of the collimated SR beam  reaching the IC, \emph{i.e.}, the IC must be able to count the SR photons along with position information. According to Eq. ($2.1$), only the difference between total flux of SR photons, within the collimated SR beam ovals, above and below the plane of motion of electrons is of interest. Figure~\ref{fig7m}A illustrates this fact by shading a portion of each SR beam oval indicating a lower flux of SR photons in that region compared to the unshaded region of the oval. With the help of an IC using central split-plane cathode, as illustrated in Figure~\ref{fig7m}, can pick out such up-down asymmetry and the signal $I_1^{L,R} - I_2^{L,R}$ will then be directly proportional to electron beam polarization. But by using the signal $S1 - S2$, also directly proportional to the electron beam polarization, one could eliminate issues introduced by vertical electron beam motion \cite{[3]}. The fact that the signal $S1 - S2$ is only directly proportional to the electron beam polarization makes this setup, at best, a good differential polarimeter.

\begin{wrapfigure}{h!}{0.5\textwidth}
\vspace{-3mm}
\centering
    \includegraphics[scale=.2]{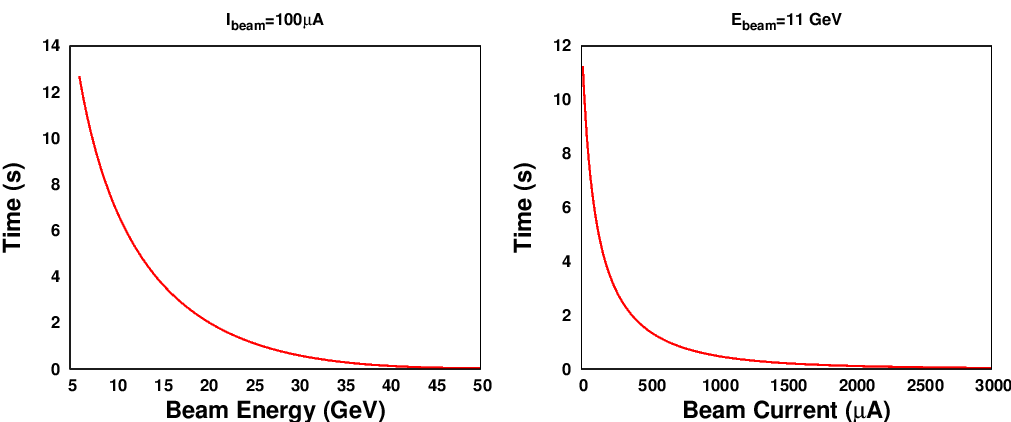}
\caption[]{(Left-Right) A. A plot of response time of the IC as a function of electron beam energy held at a fixed electron beam current of $100\mu A$.; B. A plot of response time of the IC as a function of electron beam current held at a fixed electron beam energy of $11GeV$ }
\label{fig9}
\vspace{-3mm}
\end{wrapfigure}

\begin{wrapfigure}{h!}{0.45\textwidth}
\vspace{-15mm}
\centering
    \includegraphics[scale=.25]{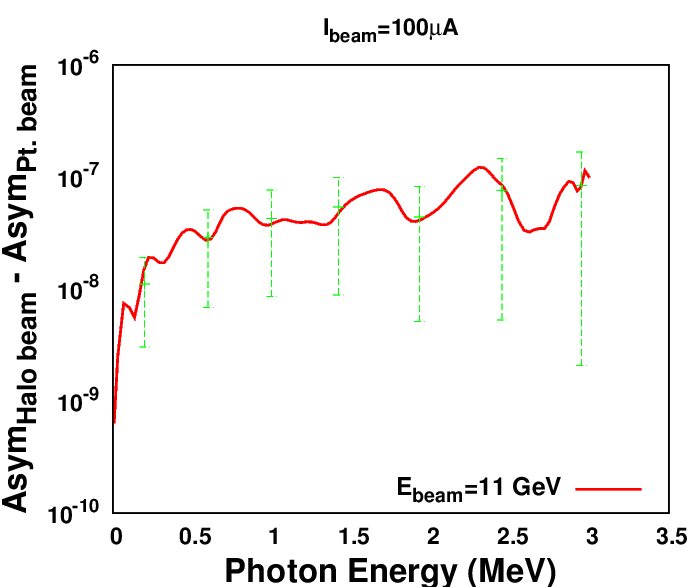}
\caption[]{A plot showing the difference in asymmetry introduced by the addition of a beam halo generated using the Geant4 simulation.}
\label{fig10}
\end{wrapfigure}

To study the IC response and behavior, a full fledged Geant4 \cite{[5]} simulation of the entire spin-light apparatus was built and its validation is presented in Ref.~\cite{[4]}. The SR beam ovals indicated in Figure~\ref{fig7m}A was regenerated in Geant4 using collimator slit width of $100\mu m$ and is presented in Figure~\ref{fig7}. Given that the SR cone has a $1mrad$ angular spread, the ovals in Figure~\ref{fig7} has a rough semi-major axis of $1cm$ with the distance between the wiggler magnets and the IC being about $10m$. IC characteristic response time could be defined as time required to reach $1/e$ of the maximum current ($I_{1,2}^{L,R}$) that the IC can provide before the IC saturates, when a collimated SR beam, produced by well defined electron beam, is incident on the IC. Figure~\ref{fig9} plots this response time for a ionization chamber with dimensions $1m\mbox{(z-axis)} \times 10cm \times 10cm$ filled with Xenon gas at $1ATM$ pressure. While it is usual to observe a quick drop in response time of IC as electron beam energy increases, it is important to note that the response time falls quickly even as electron beam current increases. Even though one would naively expect a linear decay in response time with increase in electron beam current, one must note that that SR beam power spectra changes significantly with change in electron beam current (as illustrated in Figure~\ref{fig3}C \& D) leading to  quick decay in response time with increasing electron beam current. Finally, with the help of Geant4, the effect of addition of a realistic beam halo, with a one - thousandth peak amplitude as compares to the peak amplitude of the cylindrical Gaussian beam, can be studied. In Figure~\ref{fig10}, the difference in asymmetry with and without the above mentioned halo is presented.

\vspace{-3mm}
\section{Conclusion}
\vspace{-1mm}

The spin-light polarimetry technique presents a novel method which is not just capable of achieving a high degree of precision, but is also a non - invasive and continuous technique. Such a polarimeter could ideally be implemented as a differential polarimeter which could easily be operated in series with Compton or M\o ller polarimeters. The small response time of the IC indicates that it will have to be operated with a duty cycle of less than $100\%$ in order to allow it to recover from saturation. This adds overages to the time required to obtain $1\%$ statistics but owing to small recovery times (similar to response time) and small statistical run time requirement, this is not debilitating. Ref.\cite{[3]}, tabulates a number of systematic uncertainty sources. To that list, one might add a maximum halo contribution of about $0.1\%$ owing to a residual asymmetry of about $10^{-7}$ compared to the halo less assymetry of $10^{-4}$. To precisely list all the sources of errors is a nevertheless a challenge.

\end{document}